\documentclass[doublecol]{epl2}
\usepackage{graphicx,graphics,color}% Include figure files
\usepackage{epstopdf}
\usepackage{color}

\title{Optical Properties of Organometallic Perovskite:
An {\it ab initio} Study using Relativistic {\it GW} Correction and Bethe-Salpeter Equation}
\author{Towfiq Ahmed\inst{1}\thanks{Corresponding author: \email{atowfiq@lanl.gov}}  \and C. La-o-vorakiat\inst{2} \and T. Salim\inst{3}\and Y. M. Lam\inst{3}
\and Elbert E. M. Chia\inst{2}\thanks{Corresponding author: \email{ElbertChia@ntu.edu.sg}} \and  Jian-Xin Zhu\inst{1,4}
\thanks{Corresponding author: \email{jxzhu@lanl.gov}}
}
\institute{
\inst{1} Theoretical Division,  Los Alamos National Laboratory,
Los Alamos, New Mexico 87545, USA \\
\inst{2} Division of Physics and Applied Physics, School of Physical and
Mathematical Sciences, Nanyang Technological University, 637371 Singapore\\
\inst{3} School of Materials Science and Engineering, Nanyang Technological University, 639798 Singapore\\
\inst{4} Center for Integrated Nanotechnologies, Los Alamos
National Laboratory, Los Alamos, New Mexico 87545, USA
}


\abstract{
In the development of highly efficient photovoltaic cells,
solid perovskite systems
have demonstrated unprecedented promise, with the figure of merit exceeding
nineteen percent of efficiency.  In this paper,
we investigate the optical and vibrational properties of organometallic
cubic perovskite CH$_3$NH$_3$PbI$_3$ using first-principles calculations. For accurate theoretical
description, we go beyond conventional density functional theory (DFT),
and calculated optical conductivity using relativist quasi-particle ($GW$)
correction. Incorporating these many-body effects, we further solve
Bethe-Salpeter equations (BSE) for excitons, and  found enhanced optical conductivity near the
gap edge.  Due to the presence of
organic methylammonium cations near the center of the perovskite cell, the
system is sensitive to low energy vibrational modes.  We estimate the
phonon modes of CH$_3$NH$_3$PbI$_3$ using small displacement approach, and
further calculate the infrared absorption (IR)
spectra. Qualitatively, our calculations of low-energy phonon
frequencies are in good agreement with our terahertz measurements.
Therefore, for both energy scales (around 1.5 eV and 0-20 meV), our calculations
reveal the importance of many-body effects and their contributions to the
desirable optical properties in the cubic
organometallic perovskites system.
}

\begin{document}

\maketitle

\section{Introduction}
Current progress in hybrid perovskite systems have shown significant
promise in
developing efficient yet low cost
photovoltaics~\cite{MALoi:2013}. Since the discovery of organometallic perovskite
CH$_3$NH$_3$PbX$_3$ (X=Cl, Br, I) by Kojima {\it {et al.}}~\cite{kojima},
there have been several
experimental investigations on different phases of these systems. Higher
power conversion efficiencies (PCE) have been reported by several
groups, which exceeds 12\% conversion rate~\cite{Burschka2013,MLiu2013}.
Recently Heo {\it {et al.}}~\cite{heo} have found 12.3\% PCE in
CH$_3$NH$_3$PbI$_3$ on
TiO$_2$ substrate.
The thin films of CH$_3$NH$_3$PbI$_3$ have also been reported~\cite{Snaith,Xing,sun} to have a very
high diffusion (about 100 nm) length for both electrons and holes,  which makes
them excellent candidates for optoelectronic devices. This leads one
to believe that excitons may play an important role for energy transfer mechanism in
these systems. In addition, there have also been several experimental~\cite{stab_exp_1,stab_exp_2} and
theoretical\cite{stab_th_1,stab_th_2} studies
on the structural stability in hybrid perovskite CH$_3$NH$_3$PbI$_3$.
The high sensitivity of fundamental band gap to various stable
crystal-structure  phases of CH$_3$NH$_3$PbI$_3$ has been reported~\cite{stab_th_1,stab_th_2}, suggesting the importance of the structure stability in
 both
solar-cell applications and optoelectronic devices.  More recently, it has also been predicted~\cite{AAmat:2014} that the cation-induced structure variability can promote strikingly different electronic and optical properties.

In this Letter,
we focus on the excitonic effects on the optical conductivity of cubic CH$_3$NH$_3$PbI$_3$ using
first-principles methods.
The excitons are electron-hole pairs, a phenomenon, which is often found in
wide band-gap
semiconducting systems. These are associated with bound states formed between
the excited
electrons and the remaining holes.
To capture these excitonic effects, more specifically the Coulomb
interaction between the excited electrons and the remaining holes, we
go beyond the random-phase approximation (RPA) and DFT-based single particle theory,
and solve the equation of motion for
two-particle response functions, formally known as Bethe-Salpeter equations
(BSE)~\cite{onida,strinati_2,steven,albrecht}.
%The contribution from excitonic states are
%incorporated in the
%dielectric functions by solving the Bethe-Salpeter equations as implemented in the YAMBO code.\cite{yambo,yambo_code} Such excitonic effects are best
%accounted for when the quasi-particle corrections are performed
%on top of single particle DFT calculations.
%Here, we have calculated the BSE resolved dielectric function in
%addition to the
%single shot $GW$ self-energy correction.
Due to the presence of significant spin-orbit  coupling (SOC) in Pb ions,
we also perform calculation using our own generated full-relativistic pseudo-potential,  and compare the results with non-relativistic (no SO coupling)
calculations.
We observe that both SOC and  BSE with {\em GW}-correction play a significant role in optical properties of
CH$_3$NH$_3$PbI$_3$ system. We find that the obtained exciton has about 0.153 eV of binding energy for a band gap of 1.48 eV.
Our results may provide an alternative interpretation of optical response as measured by photoluminescence spectroscopy~\cite{yamada}.
We also calculate the vibrational modes and  infrared absorption (IR) spectra based on the ionically relaxed structure of cubic CH$_3$NH$_3$PbI$_3$.
The modes in the low frequency region are in a qualitative agreement with dominant peak features observed in our THz conductivity measurement on a CH$_3$NH$_3$PbI$_3$ thin film.

\begin{figure}
 \includegraphics[scale=0.24,angle=0]{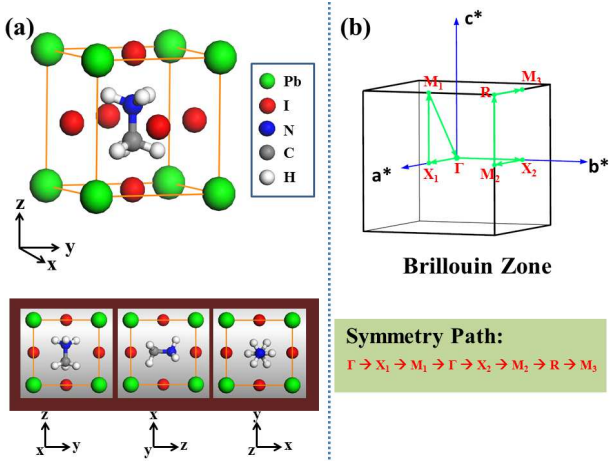}
  \caption
   {(Color online)
Ionic relaxed crystal structure and Brillouin Zone for the organometallic cubic
perovskite CH$_3$NH$_3$PbI$_3$; (a) The final optimized
configuration of the methylammonium cation in side the cubic cell. Lower panel
shows
several cross-sectional view along $\bf{x}$, $\bf{y}$ and, $\bf{z}$ axis.
(b) The $\mathbf{k}$-paths along high symmetry points, along which the band structure is displayed.
}\label{f1}
\end{figure}

\section{Computational Details and Experimental Method}
Experimental characterization of the crystal structure of hybrid perovskites are difficult. The X-ray diffraction on high-quality CH$_3$NH$_3$PbI$_3$  crystals has identified cubic, tetragonal, and orthorhombic phases; while the transmission electron microscopy has suggested a pseudo-cubic phase to be consistent with the variability in the octahedral tilting~\cite{synthesis}.  Throughout the paper, the cubic phase is considered by neglecting small distortions as in Ref.~\cite{stab_th_1}.
We perform the first-principles calculations in several steps.
We start with a DFT-based structural optimization by relaxing
the internal coordinate of
the ions in the cubic CH$_3$NH$_3$PbI$_3$ unit cells while keeping the
lattice constant and cell shape fixed (see Fig.~\ref{f1}(a)).
The force on each atom was optimized within 5 meV/\AA. For these calculations,
we used fixed value of $a=b=c=6.26\; \AA$  for the lattice constants,
which are obtained from the powder diffraction experiment~\cite{stab_th_1,synthesis}.

For the above DFT step, we use {\it ab initio} package VASP~\cite{vasp_main}
and choose
ultra-soft pseudo potentials~\cite{vasp_pseudo_1,vasp_pseudo_2} with PBE exchange correlation
functional~\cite{vasp_pbe_1,vasp_pbe_2}.
The first Brillouin zone (BZ) is sampled with $4 \times 4 \times 4$ $k$-points using
the Monkhorst Pack grid. Once the ionic and electronic relaxations have been
achieved, we employ the small-displacement approach to calculate the
vibrational
eigen-frequencies and eigenvectors. Linear response method is used to
calculate the Born effective charge tensors. The infrared-active modes are
then obtained by the corresponding oscillator strength~\cite{IR_oscillator}:
\begin{equation}
f(\omega_n)=\sum_{\alpha}
\left| \sum_{i\beta} Z^{*}_{\alpha \beta}(i)v_{\beta}
(i|\omega_n) \right|^2,
\end{equation}
where $Z^{*}_{\alpha \beta}(i)$ is the effective charge tensor of $i$-th atom,
$v_{\beta}(i|\omega_n)$ is the vibrational eigenvector for $i$-th atom and
$n$-th eigen-frequency $\omega_n$, and $\alpha$, $\beta$ are the three
components of the cartesian axis.
Using these oscillator strengths, we simulate the experimentally observable
IR spectra based on a simple analytical expression~\cite{IR_lorentz}:
\begin{equation}
I(\omega) = \sum_n f(\omega_n)\delta_m(\omega-\omega_n)/\sum_n f(\omega_n),
\end{equation}
where $\delta_m(\omega)$ are delta like functions defined by:
\begin{equation}
\delta_m(\omega)=\frac{m}{\pi} \frac{1}{1+m^2\omega^2}.
\end{equation}
The broadening of the Lorentzian peaks are adjusted to have the best agreement with
experimental data (see Fig.~\ref{f4}).

\begin{figure}
 \includegraphics[scale=0.24,angle=0]{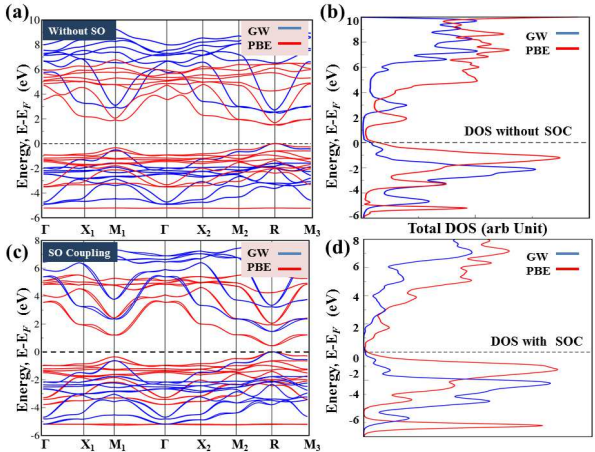}
  \caption
   {(Color online)
DFT band structure and total DOS with $GW$ (solid blue curve) and without $GW$
(solid red curve)
correction. Top right and left panels ((a) and (b)) show the total DOS and
band structure
without the SOC. Bottom panels ((c) and (d)) show results
with the SOC. Horizontal dashed lines show the location of Fermi energy.
}\label{f2}
\end{figure}

For the excitonic contribution to the optical conductivity, we perform a single
shot $GW$
correction on top of our DFT calculations,  and finally solve the BSE as implemented in the code YAMBO~\cite{yambo}.
In $\Sigma(\omega)=G^0(\omega)W(\omega)$ self-energy formalism,
$G^0$ stands for the non-interacting Green's function,
where $W$ is the screened Coulomb interaction,
$W(\bf{r},\bf{r'},\omega)=\epsilon^{-1}(\bf{r},\bf{r'},\omega)
V(\bf{r},\bf{r'})$. Here, $V(\bf{r},\bf{r'})$ is the bare Coulomb
interaction.
The frequency dependent dielectric function $\epsilon(\bf{r},\bf{r'},\omega)$
is
calculated from the response function $\chi(\omega)$ as~\cite{hedin69,Hedin_1999, Hedin_PrB99, hedin-lundqvist69}:
\begin{equation}
\epsilon^{-1}_{\bf{G},\bf{G'}}=\delta_{\bf{G},\bf{G'}}+V(\bf{q}+\bf{G})\chi_{\bf{G},\bf{G'}}(\bf{q},\omega),
\end{equation}
where within the random phase approximation (RPA)\cite{hedin69,onida}, the response function
$\chi_{\bf{G},\bf{G'}}(\bf{q},\omega)$ can be calculated from non-interacting
$\chi^0$ and Green's function $G^0$. The quantity $\bf{q}=\bf{k}-\bf{k'}$
is momentum transfer and $\bf{G}$ is the reciprocal lattice vectors.
This method was first proposed by Hedin~\cite{hedin_1}, and further
details can be
found in Refs.~\cite{aryasetiawan_98, strinati_1, albur_2000, inkson_84, Kotani2008,
Hedin_1999}.  It has been found that the inclusion of SOC in the GW calculations is also important in 5$f$-electron
materials~\cite{dirac-GW,towfiq:2014}.

This $GW$ method is a many-body perturbation  technique which
accounts for
missing dynamical correlation in the DFT.
This GW-correction method has been used to successfully predict accurate band gaps for
various narrow-gap semiconductors. Therefore, in CH$_3$NH$_3$PbI$_3$
system under consideration, we perform $GW$ correction on top of the DFT calculations.
Convergence of GW calculations with respect to various paramaters are often nontrivial and
necessary to predict the accurate band gap in the wide gap semiconductors and insulators. In this work, we have
used one shot GW using the Godby-Needs (GN) plasmon-pole model~\cite{godby} as implemented in Yambo. Previously
such GN model
was demonstrated to agree very well with the model free GW self-energy for predicting band gap in other
systems~\cite{ZnO_pp}. Using this framework within Yambo, we have achieved the convergence of our GW
band-structure calculations
with 200 bands and 36 momentum
transfer $\mathbf{q}$ points in the irreducible Brillouin zone. Furthermore, we have also converged our calculations with
respect to polarization matrix dimension and with 200 bands in the evaluation of polarization function.
Our GW convergence is also performed with respect to the wave function cutoff energy at the DFT level, for which
we have found 70 Ry is sufficient for our system of interest.

\begin{largetable}
\caption{ Fundamental band gap and lattice parameters in cubic
CH$_3$NH$_3$PbI$_3$.}
\centering
\begin{tabular*}{0.58\textwidth}{@{\extracolsep{\fill}} l  c  c  c  c}
\hline
      & This Work & Other Work & Expt. \\[0.5ex]
\hline\hline \\[-0.3ex]
a=b=c  & 6.26 \AA & -- & 6.26 \AA\cite{synthesis}  \\
$\alpha=\beta=\gamma$ &  90$^{\circ}$  & --  \\
gap (PBE) & 1.51 eV  & 1.38 eV\cite{mark_van} & --  \\
gap (PBE+$GW$) & 2.53 eV  &  -- & -- \\
gap (PBE+SO) & 0.46 eV  & 0.53 eV\cite{mark_van} &  --  \\
gap (PBE+SO+$GW$) & 1.48 eV & 1.27 eV\cite{mark_van} &  1.61 eV~\cite{yamada}  \\
gap (PBE+SO+sc-$GW$) & -- & 1.67 eV\cite{mark_van} &  1.61 eV~\cite{yamada}  \\
\hline
\end{tabular*}
\label{table}
\end{largetable}

However, for systems with larger band gap, the RPA is no longer adequate, and one has to
incorporate electron-hole (e-h) correlation in the response function $\chi$. This is the
Bethe-Salpeter equation (BSE), where one calculates the response function in terms of the
noninteracting two particle e-h Green's function. The details of this
formalism can be found in Refs.~\cite{yambo,onida,strinati_2}. Alternatively, one can also solve the BSE
by reducing the problem to a two-particle e-h Hamiltonian~\cite{onida}. The eigenvectors of this Hamiltonian
are the excitonic states $\left| \lambda \right>$ with $E_{\lambda}$ being the corresponding  excitonic binding energies  within the Tamm-Dancoff\cite{fetter}
approximation, in which only positive energies are considered after diagonalizing the non-Hermitian e-h Hamiltonian~\cite{yambo}.
For narrow-gap semiconductors, the effect of BSE is usually negligible; while for wide gap semiconductors,
$E_\lambda$ are often within the gap and can thus have significant contribution to the optical properties.
%\textcolor{red}{ Our BSE convergence are also tested on equal footing with GW calculations. We have reached
%high level of convergence for excitonic states using 36 $q$ points and 100 bands (30 of which are occupied).
%}
%In our calculations for CH$_3$NH$_3$PbI$_3$, the $GW$ corrected band gap is
%$\sim$ 1.85 eV
%including the SO coupling,
%and therefore, we calculate the excitonic contributions to optical conductivity solving BSE as implemented in YAMBO.
The YAMBO program uses single-particle quantities based on the DFT
calculations within the  Quantum Espresso (QE)~\cite{QE-2009}. Therefore, we export the VASP optimized
structure to QE and, through the ionic force calculation, verify the optimized structure from the VASP. The relaxed
structure is then used to calculate the electronic band-structure and total
density of states.
We use
norm-conserving PBE based pseudopotentials in QE.
To account for the SO coupling,
we generated fully relativistic norm-conserving PBE pseudopotential by including the 5d semicore electrons as the
valence states for the Pb ions. For both cases (with and
without SO coupling), QE generates DFT band-structure, which are then
renormalized with $GW$ corrections using YAMBO. Finally,
excitonic contributions to the optical properties are incorporated on top of
DFT and DFT+$GW$ calculations including SO coupling. Convergence test on our BSE calculations were performed on equal footing with
GW calculations. We reached high level of convergence in our estimation of excitonic states by using 80 bands (60 of which are unoccupied) and 36 $\mathbf{q}$ points.
Further convergence was tested with respect to polarization bands and matrix dimension. Similar convergence criteria were also reported in previous
work~\cite{haibin}.

The THz transmission spectra of the CH$_3$NH$_3$PbI$_3$ perovskite sample was obtained using Teraview TPS3000 THz time-domain spectrometer, coupled with a Janis ST-100-FTIR cryostat. The perovskite film of thickness 230 nm is deposited on a $z$-cut quartz substrate.
A bare substrate is used for the reference run. By fitting the spectra with the thin-film-on-substrate
transmission expression, one obtains the complex THz conductivity of the sample.
The experimental details are described in Ref.~\cite{XZou:2013}.
Data taken at 300 K and 20 K are presented in the present work.

\section{Results}
All our calculations presented in the Letter are based on highly optimized structure of
cubic CH$_3$NH$_3$PbI$_3$ system. The ionic relaxation are performed for internal
coordinates while we kept the lattice parameters fixed. The final configurations of the
atoms are shown in Fig.~\ref{f1}(a). In the lower panel of Fig.~\ref{f1}(a), we show $\bf{x}$, $\bf{y}$,
and $\bf{z}$ projected cross-sectional view, which shows the final orientation of methylammonium cation
inside the cubic cell. For the band structure calculations, we use
high symmetry $\mathbf{k}$-paths inside the first Brillouin zone as shown in Fig.~\ref{f1}(b).

\subsection{SOC and QP correction}

We have systematically studied the effect of SOC as well as $GW$ self-energy corrections
on the fundamental band gap of cubic CH$_3$NH$_3$PbI$_3$. The PBE-DFT calculations without
SOC predicts 1.5 eV gap (solid red line in Fig.~\ref{f2}(a)) while $GW$ correction (without SOC)
estimates the gap to be 2.5 eV (solid blue line in Fig.~\ref{f2}(a)).
The inclusion of SOC reduced the band gap to be 0.46 eV for DFT (solid red line in Fig.~\ref{f2}(c))
while the $GW$ corrections corrects the band gap up to 1.48 eV (solid blue
curve in Fig.~\ref{f2}(c)). The corresponding total density of states (DOS) are shown in Fig.~\ref{f2}(b) and (d).
The experimental band gap for this system was found to be 1.61 eV at room temperature~\cite{yamada}.
Therefore, we see that the inclusion of SOC underestimates the band gap by more than one eV, while the $GW$ correction
without SOC overestimates
the gap by about 0.9 eV.  Our systematic study reveals that the best agreement with experiment is reached when we include both the SOC and
 $GW$ self-energy
correction.
%We notice that the the experimental value  of the band gap is obtained at room temperature, where the thermal effect could reduce the band gap as compared to that in the ground state.
All our calculated values of band gap at various conditions are listed in Table I.
We point out that our finding of the effects of SOC and self-energy correction on band gap agree well with earlier results reported by Brivio {\it {et al.}}~\cite{mark_van} for the cubic phase and by Amat {\em et al.}~\cite{amant} for the tetragonal phase of the perovskite.
For orthorhombic and tetragonal phase, Zhu {\it {et al.}}~\cite{haibin} have reported
the band gap correction to be 1.69 eV and 1.57 eV correspondingly using GW and SOC. Umari {\it {et al.}}~\cite{umari} have also reported
such band gap around 1.67 eV while the experimental band gap in such systems was determined to be 1.6 eV. We ascribe such a small discrepancy to the fact that various plasmon models have been used for the frequency dependence of the dielectric matrix.
\begin{figure}
 \includegraphics[width=1.0\linewidth,clip]{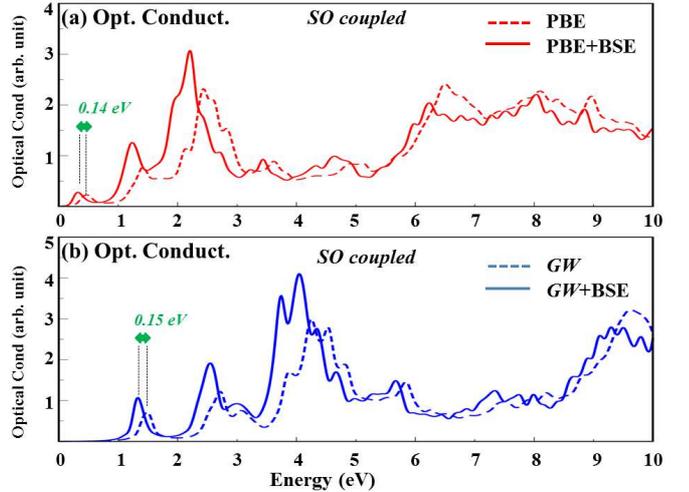}
  \caption
   {(Color online)
(a) Optical conductivity with BSE (solid red) and without BSE (dashed red)
corrections using the DFT calculations;
(b) Optical conductivity with BSE (solid blue) and without BSE (dashed blue)
using the DFT plus $GW$ self-energy correction. The arrows in both panels (top
and bottom) point the excitonic contribution to the optical conductivity.
All calculations include SOC.
}\label{f3}
\end{figure}

\subsection{Excitonic Effect}

In the presence of SOC, we calculated the dielectric function and optical conductivity by solving BSE  without (Fig.~\ref{f3}(a)) and with (Fig.~\ref{f3}(b))  the self-energy
correction. The peaks in optical conductivity represent the electronic excitations from valence band to conduction band satisfying
 the optical selection rule~\cite{select_optic, select_exciton}.
Therefore, the fundamental
gap can be represented by the location of the first peak in optical conductivity spectrum. By incorporating the
excitons through the solutions of BSE, we find enhanced peak appearing below the gap edge. These peaks are indicated by the arrows in Fig.~\ref{f3}.
We also find that the BSE solutions including $GW$ self-energy correction, as shown in Fig.~\ref{f3}(b), give rise to more pronounced
excitonic contribution than without $GW$ correction (Fig.~\ref{f3}(a)). By comparing the location of this shifted excitonic peak with
the band gap edge, we are able to deduce that the exciton binding energy is  about 0.153 eV. Interestingly, we observed the excitonic binding energy for the cubic phase is larger than for the tetragonal or orthorhombic phase reported earlier in literature~\cite{haibin}. Experimentally, it is important to find out the implication of a higher binding energy of excitons to the long diffusion length observed in perovskite thin films or heterostructures, where the strain effect from the substrate or sandwiching materials may play an important role.

\begin{figure}
 \includegraphics[scale=0.24,angle=0]{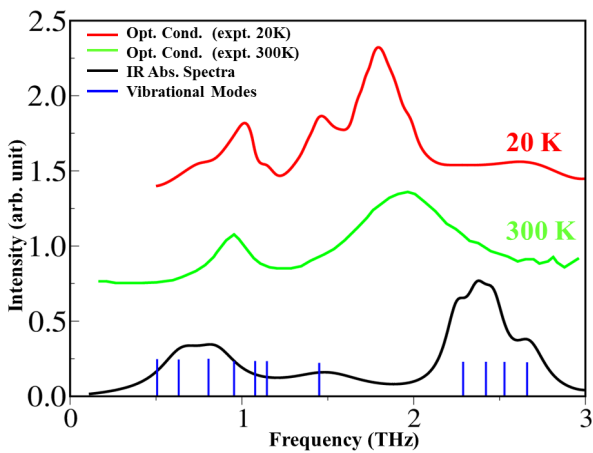}
  \caption
  {(Color online) Vibration mode and optical spectra.
The top curves (solid red and green) represent experimental data for terahertz conductivity on a CH$_3$NH$_3$PbI$_3$ thin film.
The bottom curve (black) is the calculated
infrared absorption spectrum. Vertical blue lines are the location of vibrational mode frequencies.
}\label{f4}
\end{figure}

\subsection{Vibrational Modes vs. Measurements}
The presence of flexible methylammonium cation at the center of the rigid perovskite cage implies that the crystal structure of this organometallic compound is softer than that of transition metal oxide perovskite with the former having expanded lattice constants. This observation motivates us to investigate the vibrational modes, which
can be probed by low-frequency  spectroscopy such as Raman, IR, or THz conductivity.
For this purpose, the vibrational mode energy and oscillator strength are calculated. The IR spectra are simulated with a Lorentzian fitting. The results are shown in Fig.~\ref{f4}, with the vertical blue lines denoting the positions of vibrational mode frequencies. We see that our calculated IR absorption spectra, as represented by the solid black curve, reproduces qualitatively similar peak features observed in experimental data for the THz conductivity (red and green lines), in the low frequency region (0-3 THz).
The computed spectra is slightly stretched  in the frequency range.
Such stretch has  also been reported
by earlier {\it{ab initio}} calculations of Raman spectra for CH$_3$NH$_3$PbI$_3$~\cite{raman}.
Our  THz conductivity measurement demonstrates the possible energy and charge transfer mechanism in cubic CH$_3$NH$_3$PbI$_3$
through the low-energy vibrational phonon modes.

\section{Conclusion}

In summary, we have demonstrated the effect of excitons and vibrational modes on the optical properties of cubic organometallic perovskite CH$_3$NH$_3$PbI$_3$.
All our calculations have been based on internally relaxed stable ionic structure with the experimental lattice parameters. Our study has uncovered  the importance
of  both the SOC and quasi-particle self-energy correction, which conspire to reproduce the experimentally observed fundamental band gap. In this wide-gap semiconductor, we have found significant excitonic contribution to the optical conductivity at about $\sim$ 1.5 eV using first-principles BSE calculations. In addition, we have also performed first-principles simulations of the vibrational modes due to the presence of flexible methylammonium cations at the center of the cubic perovskite cage. We have shown that the signature of the modes can be revealed in the IR spectra, with the profile very similar to the THz conductivity we have measured on a CH$_3$NH$_3$PbI$_3$ thin film.
Therefore,  our first-principles calculations have predicted that the features of the excitation spectra must  originate from the strong SOC, many-body effects, and low-energy vibrational modes.
Our systematic first-principles study, for the first time, has benchmarked  the various competing effects, which must be taken into account to understand and utilize the optoelectronic properties of CH$_3$NH$_3$PbI$_3$ to enhance its solar-cell applications.

\acknowledgments
%\section{Acknowledgements}
We thank M. J. Graf, Amanda Nuekirch, Carl Greef, Davide Sangalli, A. Walsh, and B. Xiao for useful discussions.
This work was supported by U.S. DOE  at
LANL  under Contract No. DE-AC52-06NA25396 and the LANL LDRD Program
(T.A. \& J.-X.Z.), and Singapore Ministry of Education AcRF Tier 1 (RG 13/12 \& 2014-T1-001-056) and the National Research Foundation Competitive Research Programme (NRF-CRP4-2008-04) (E.E.M.C.), and the Danish Council for Strategic Research (Y.M.L.). The work was supported in part by the Center for Integrated
Nanotechnologies, a U.S. DOE BES user
facility.

%\bibliographystyle{unsrt}
%\bibliography{references}

\end{document}